\documentclass [12pt]{article}
\usepackage{graphicx,amssymb,amsmath,bm}
\usepackage[symbol*]{footmisc}
\usepackage[cp1251]{inputenc}
\usepackage[english]{babel}
\usepackage{booktabs}

\makeatletter
\def\@biblabel#1{#1.\hskip-0.3em}
\makeatother

\begin{document}
\def\refname{\normalsize \centering \mdseries \bf REFERENCES}
\def\abstractname{Abstract}

\begin{center}
{\large \bf Polarization of nucleons in the deuteron stripping \\ reaction on nuclei}
\end{center}

\begin{center}
\bf\text{V.~I.~Kovalchuk}
\end{center}

\begin{center}
\small
\textit{Department of Physics, Taras Shevchenko National University, Kiev 01033, Ukraine}
\end{center}

\begin{abstract}
A general analytical expression has been obtained in the diffraction approximation for the
polarization of nucleons arising in the deuteron stripping reaction on nuclei at intermediate
energies of the incident particles. A tabulated density distribution of the target nucleus
and a realistic wave function of the deuteron with correct asymptotic limits at large
internucleon distances were used in the calculations. The nucleon-nucleus phase shifts were
calculated in the Glauber formalism using the double folding potential. The calculated angular
dependencies for the vector analyzing power $A_{y}$ of the reaction
$^{3}\text{He}({\textit{\textbf{d}}},p)^{4}\text{He}$ are found to be in satisfactory agreement
with the corresponding experimental data.
\end{abstract}

\bigskip
\begin{center}
\bf{1.~INTRODUCTION}
\end{center}
\smallskip

Nuclear reactions involving deuterons are widely used to examine the properties of the nuclear interaction
and nuclear structure. The importance of such reactions in experimental nuclear physics is due both to the
relative simplicity of obtaining monochromatic deuteron beams with precisely calibrated polarization and to
the higher yield of \textit{dA}-reactions in comparison with reactions induced by other particles.  By virtue of
the smallness of the binding energy of the deuteron, when a deuteron collides with a nucleus inelastic
processes are most likely to occur: breakup of the deuteron in the Coulomb field of the target nucleus
(mainly at low energies) and deuteron stripping, which occurs when one of the nucleons making up the deuteron
is absorbed by the target and the other one is liberated as a reaction product. At intermediate energies the
stripping reaction is the result of direct interaction (nucleon capture by the nucleus), and its differential
cross section is characterized by a narrow peak at nucleon emission angles ${\Theta\ll1}$.

A formalism for the inclusive deuteron stripping reaction at intermediate energies was first proposed by
Serber~\cite{1} and was then developed further by Akhiezer and Sitenko~\cite{2,3}.  In~\cite{4} it was shown
that the quintuple integral in the formula for the cross section of the reaction~\cite{3} can be calculated
analytically if the integrand functions are expanded in a series over a Gaussian basis.  In this approach,
the scattering phase shifts were calculated in the Glauber formalism using the double folding potential~\cite{5}
and tabulated distributions of the nucleon densities of the target nuclei~\cite{6}, so that the final result
(cross section) depends on a single parameter -- the normalization parameter of the imaginary part of the
nucleon-nucleus potential. The present work considers the more complex problem of the polarization of the
nucleon reaction products of deuteron stripping. The nature of the origin of such polarization, for example,
on protons, corresponds to the different probability of absorption of a neutron for the two possible orientations
of its spin~\cite{7}.  Since the spins of the nucleons in a deuteron are parallel, the direction of the
polarization of the escaping protons will be dictated by the orientation of the spin of the neutron, which has
a greater probability of absorption. It is clear that the yield of polarized protons in the stripping reaction
will be increased if the incident deuterons are polarized, and that the direction of their polarization will
coincide with the direction of the protons arising during stripping.

\bigskip
\begin{center}
\bf{2.~FORMALISM}
\end{center}

As our targets, we choose light and intermediate nuclei since for such nuclei and intermediate energies of
the deuterons in this problem the Coulomb interaction can be neglected. The polarization of the nucleons
produced by a stripping reaction involving unpolarized deuterons is expressed in terms of the density matrix
$\rho$ as follows~\cite{8}:
\begin{equation}
P={\rm{Tr}}\,(\sigma\rho)/{\rm{Tr}}\,\rho\,,
\label{eq1}
\end{equation}
where $\sigma$ is the nucleon spin operator (the Pauli matrices).

Let the deuteron move in the positive direction of the $z$-axis of a Cartesian coordinate system defined such
that the $xy$-plane is the plane of the impact parameter. We denote the proton entering into the composition
of the deuteron by the subscript 1, and the neutron, by the subscript 2. In the diffraction approximation,
the general expression for the density matrix of the stripping reaction has the form~\cite{3}
\begin{equation}
\rho({\textit{\textbf{k}}}_{1})=\int d{\textit{\textbf{b}}}_{2}(1-|1-\Omega_{2}|^{\,2}
)a_{{\textit{\textbf{k}}}_{1}}({\textit{\textbf{r}}}_{2})a_{{\textit{\textbf{k}}}_{1}}^{\dagger}
({\textit{\textbf{r}}}_{2}),
\label{eq2}
\end{equation}
where ${\textit{\textbf{b}}}_{2}$ is the impact parameter vector of the neutron and $\Omega_{2}$ is the
corresponding profile function. The quantity
\begin{equation}
a_{{\textit{\textbf{k}}}_{1}}({\textit{\textbf{r}}}_{2})=(2\pi)^{-3/2}\int d{\textit{\textbf{r}}}_{1}
\exp(-i{\textit{\textbf{k}}}_{1}{\textit{\textbf{r}}}_{1})(1-\Omega_{1})\varphi_{0}({\textit{\textbf{r}}}),
\quad{\textit{\textbf{r}}}={\textit{\textbf{r}}}_{1}-{\textit{\textbf{r}}}_{2},
\label{eq3}
\end{equation}
is the probability amplitude that the proton will have momentum ${\textit{\textbf{k}}}_{1}$ and the
neutron will be found at the point ${\textit{\textbf{r}}}_{2}$, and $\varphi_{0}({\textit{\textbf{r}}})$
is the wave function of the deuteron.

The neutron profile function $\Omega_{2}$ in expression (\ref{eq2}) contains only a radial part,
i.e., $\Omega_{2}(b_{2})=\omega_{2}(b_{2})$, but the proton profile function also depends on the
spin~\cite{9}
\begin{equation}
{\Omega}_{1}({\textit{\textbf{r}}}_{1})=\omega_{1}(b_{1})\{1+\gamma_{1}
\exp(i\delta_{1}){\sigma}(({\textit{\textbf{k}}}/2)\times\partial/\partial{\textit{\textbf{r}}}_{1})\},
\label{eq4}
\end{equation}

Here  $b_{1}$, $\gamma_{1}$, and $\delta_{1}$ are, respectively, the impact parameter, a constant,
and the phase shift of the spin-orbit interaction of the proton with the nucleus, and ${\textit{\textbf{k}}}$
is the momentum vector of the incident deuteron.

If Gaussian functions stand under the integral sign in Eq.(\ref{eq2}), then the density matrix can be
calculated in explicit form. Without loss of generality, we expand the functions $\omega_{i}(b_{i})$
$\mbox{(\textit{i}=1,2)}$ and $\varphi_{0}({\textit{\textbf{r}}})$ in a Gaussian basis:
\begin{equation}
\omega_{i}(b_{i})=\sum\limits_{j=1}^{N}{\alpha_{ij}}\exp(-b_{i}^{2}/\beta_{ij}),
\quad
\varphi_{0}({\textit{\textbf{r}}})=\sum\limits_{j=1}^{N}{c_{j}}
\exp(-d_{j}|{\textit{\textbf{r}}}_{1}-{\textit{\textbf{r}}}_{2}|^{2}),
\label{eq5}
\end{equation}
where $\beta_{ij}=R_{rms}^{2}/j$, and $R_{rms}$ is the root-mean-square radius of the target nucleus.

A analytical integration in Eq.(\ref{eq2}) with the above functions (formulas (\ref{eq5})) substituted
in formulas (\ref{eq3}) and (\ref{eq4}), followed by calculation of the traces of the matrices in the
numerator and the denominator of formula (\ref{eq1}), yields the expressions
\mbox{${\rm{Tr}}(\sigma\rho({\textit{\textbf{k}}}_{1}))={\textit{\textbf{G}}}({\textit{\textbf{k}}}_{1})$}
and \mbox{${\rm{Tr}}(\rho({\textit{\textbf{k}}}_{1}))={\textit{H}}({\textit{\textbf{k}}}_{1})$}
(see Appendix A). Thus, the polarization of the proton is given by the formula
\begin{equation}
{\textit{\textbf{P}}}({\textit{\textbf{k}}}_{1})=G({\textit{\textbf{k}}}_{1})
({\textit{\textbf{p}}}_{1}\!\times\!{{\textit{\textbf{k}}}_{1}})/
{H({\textit{\textbf{k}}}_{1})},
\label{eq6}
\end{equation}
where ${\textit{\textbf{p}}}_{1}$ is the transverse component of the momentum
\mbox{${\textit{\textbf{k}}}_{1}=\left\{{\textit{\textbf{p}}}_{1},({\textit{\textbf{k}}}/k)k_{1z}\right\}$}.
The values of the components $p_{1}$ and $k_{1z}$ of the vector ${\textit{\textbf{k}}}_{1}$ are functions
of the energy of the proton $T_{1}$ and its emission angle $\Theta_{1}$ in the laboratory coordinate
system~\cite{3}:
\begin{equation}
p_{1}=(k/2+k_{1z})\tan\Theta_{1},
\quad
k_{1z}=\sqrt{m/T}(T_{1}-T/2).
\label{eq7}
\end{equation}

Here $m$ is the nucleon mass and $T$ is the initial energy of the deuteron.

Note that the denominator in expression (\ref{eq6}) is determined by the doubly differential cross section
of the stripping reaction (with respect to the emission angle of the nucleon and its energy)~\cite{3}.

In order to find the angular distribution of the polarization, the functions $G({\textit{\textbf{k}}}_{1})$
and $H({\textit{\textbf{k}}}_{1})$ in expression (6) must be integrated with respect to the $z$-component
of the vector ${\textit{\textbf{k}}}_{1}$~\cite{3}. Expressing ${\textit{\textbf{k}}}_{1}$ in terms of the
components of the corresponding cylindrical coordinate system and taking relations (\ref{eq7}) into account,
we obtain
\begin{equation}
P(\Theta_{1})=\int_{-\infty}^{\infty}(k/2+k_{1z})^{2}G(p_{1},k_{1z})dk_{1z}/
\int_{-\infty}^{\infty}(k/2+k_{1z})^{2}H(p_{1},k_{1z})dk_{1z}.
\label{eq8}
\end{equation}
To transform the angle $\Theta_{1}$ into the corresponding quantity in the center-of-mass system (c.m.s.),
we use the formulas of relativistic kinematics given in~\cite{10}.

\bigskip
\begin{center}
\bf{3.~RESULTS AND DISCUSSION. CONCLUSIONS}
\end{center}

Current experiments measure, as a rule, not the polarization of the particles, rather the analyzing power of
the reaction~\cite{11}. These quantities are interrelated: thus, for example, for a stripping reaction with
polarized deuterons, the vector analyzing power of the reaction $A_{y}=3P$~\cite{7,11}. The formalism expounded
in the present paper was used to describe the polarization observables $A_{y}$ of the reaction
$^{3}\text{He}({\textit{\textbf{d}}},p)^{4}\text{He}$ for deuteron energies \mbox{$T=140$}, 200, and
270 MeV~\cite{12}.

The values of the parameters ${\alpha_{ij}}$ in formulas (\ref{eq5}) were calculated using nucleon-nucleus phase
shifts in the Glauber formalism~\cite{4,5} and the experimentally obtained distribution of the nucleon density
of the $^{3}\text{He}$ nucleus~\cite{6} (see Appendix B). To find the parameters $c_{j}$ and $d_{j}$ of the deuteron
wave function (formulas (\ref{eq5})), the variational problem in the basis \mbox{$N=10$} of Gaussian functions with
the K2 triplet nucleon-nucleon potential from~\cite{13} was solved separately. This wave function has the correct
asymptotic limits at large internucleon distances; moreover, it reproduces with a high degree of accuracy the
experimental values of the binding energy of the deuteron and its root-mean-square radius.

Figure 1 plots the calculated angular dependencies of the vector analyzing power of the deuteron stripping reaction
on $^{3}\text{He}$ nuclei for \mbox{$T=140$}, 200, and 270 MeV. The values of the parameters of the spin-orbit
interaction \mbox{$(\gamma_{1},\delta_{1})$} were \mbox{$(0.143,0.143)$} for \mbox{$T=140$}~MeV, \mbox{$(0.141,0.117)$}
for \mbox{$T=200$}~MeV, and \mbox{$(0.139,0.106)$} for \mbox{$T=270$}~MeV. The normalization constant of the imaginary
part of the double folding potential, $N_{I}$ (formula (B.6), Appendix B), was equal to unity in all cases.

\begin{figure}[!h]
\center
\includegraphics [scale=1.00] {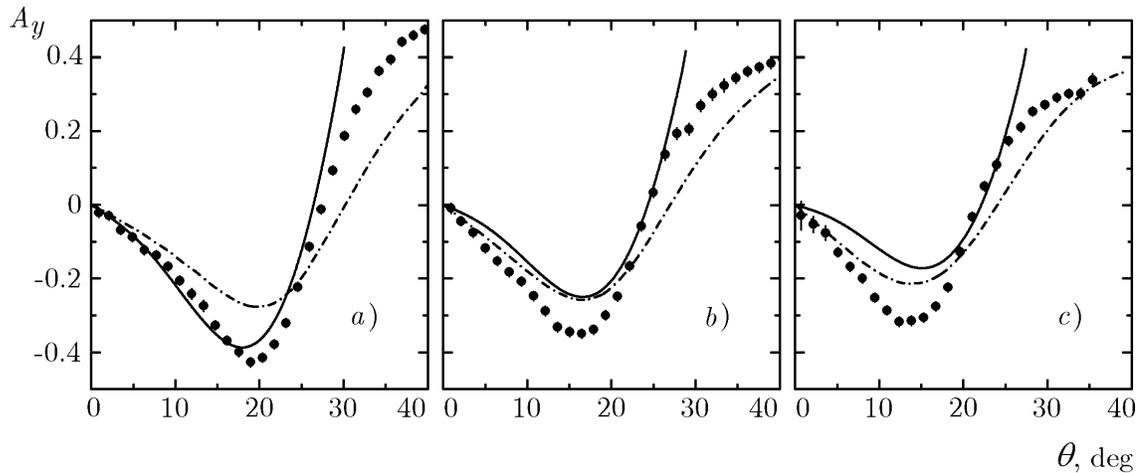}
\caption{Vector analyzing power $A_{y}$ of the reaction $^{3}\text{He}({\textit{\textbf{d}}},p)^{4}\text{He}$
as a function of the proton emission angle $\theta$ in the c.m.s. for incident deuteron energies
\mbox{$T=140$}~($a$), 200~($b$), and 270~MeV~($c$). The curves are explained in the text.
Experimental data from~\cite{12}.}
\label{fig1}
\end{figure}

The dash-dot curves in Fig.~1 correspond to the results of calculations of $A_{y}$ for the reaction of
elastic $pd$-scattering, performed in~\cite{12} on the basis of the formalism of Faddeev equations with
the realistic $NN$-potential AV18.

From an analysis of the behavior of the curves in the figure it follows that both dependencies qualitatively reproduce
the behavior of the experimental data, and the curves have a minimum at the point at which the polarization changes sign.

Thus, the formalism presented in this paper for calculating the polarization of nucleons arising in the deuteron
stripping reaction at intermediate energies allows at least a qualitative description of the corresponding experiments.
The method is based on the use of Gaussian expansions of the distributions of the nucleon density of the target nuclei
and the deuteron wave function, which made it possible to calculate the integrals analytically  and to calculate the
density matrix of the reaction and the mean value of the spin of the escaping nucleon in explicit form.

An inadequacy of the method is the relatively narrow range of angles in which agreement with experiment
is obtained (due to limitations of the diffraction nuclear model~\cite{3}). In this work we also did not
take into account the $D$-component in the deuteron wave function although the expected correction to
the vector analyzing power in this case would have amounted to several percent of the calculated value.

Overall, the proposed method can be used to analyze the corresponding experiments and determine the values
of the parameters of the spin-orbit interaction and their energy dependence. It should be noted that the
general formulas obtained in this work can be used to describe the cross sections and polarizations in
inclusive processes not only of stripping, but also of breakup of deuterons, and also of light and heavy
ions~\cite{14}. In these cases, both the angular and the energy distributions of the indicated quantities
can be calculated (in the latter case, the expressions for $G$ and $H$ appearing in expression (\ref{eq6})
must be integrated over the perpendicular components of the momentum vector of the incident particle).

\setcounter{section}{0}
\def\theequation{\Alph{section}.\arabic{equation}}
\def\thesection{\normalsize Appendix \Alph{section}}
\setcounter{equation}{0}
\section{\normalsize{Traces of the density matrix $\rho$ and the products $\sigma\rho$}}
\hspace{\parindent}
We present the result of calculation of the traces of the matrices in the numerator and the denominator
of expression (\ref{eq1}) after analytical integration in expression (\ref{eq2}) with the functions
assigned by formulas (\ref{eq5}).

The numerator in expression (\ref{eq1}) has the form
\begin{equation}
{\rm{Tr}}({\sigma}\rho({\textit{\textbf{k}}}_{1}))={\textit{\textbf{G}}}({\textit{\textbf{k}}}_{1})=
G({\textit{\textbf{k}}}_{1})({\textit{\textbf{p}}}_{1}\!\times\!{{\textit{\textbf{k}}}_{1}})\,.
\label{A1}
\end{equation}
Here ${\textit{\textbf{p}}}_{1}$ is the transverse component of the momentum of the incident nucleon:
\mbox{${\textit{\textbf{k}}}_{1}=\bigl\{{\textit{\textbf{p}}}_{1},{\textit{\textbf{k}}}_{1z}\bigr\}$},
and the quantity $G({\textit{\textbf{k}}}_{1})$ is defined as
\begin{equation}
G({\textit{\textbf{k}}}_{1})=\gamma_{1}\sin\delta_{1}\sum\limits_{s=1}^{N}
\sum\limits_{q=1}^{N}{c_{s}}{c_{q}}Z(\lambda,p_{1},k_{1z}),
\label{A2}
\end{equation}
where
\begin{equation}
Z(\lambda,p_{1},k_{1z})=2\lambda t(\lambda,k_{1z})(4z^{(1)}
(\lambda,p_{1})-z^{(2)}(\lambda,p_{1})),
\label{A3}
\end{equation}
\begin{equation}
z^{(1)}=z_{11}-z_{12},
\label{A4}
\end{equation}
\begin{equation}
z^{(2)}=4z_{21}+z_{22}.
\label{A5}
\end{equation}

The denominator in expression (\ref{eq1}) is equal to
\begin{equation}
{\rm{Tr}}\rho({\textit{\textbf{k}}}_{1})=H({\textit{\textbf{k}}}_{1})=
H_{0}({\textit{\textbf{k}}}_{1})+\gamma_{1}^{2}H_{\rm{so}}({\textit{\textbf{k}}}_{1})\,,
\label{A6}
\end{equation}
where
\begin{equation}
H_{0}({\textit{\textbf{k}}}_{1})=\sum\limits_{s=1}^{N}\sum\limits_{q=1}^{N}{c_{s}}{c_{q}}
Y(\lambda,p_{1},k_{1z}),
\label{A7}
\end{equation}
\begin{equation}
Y(\lambda,p_{1},k_{1z})=t(\lambda,k_{1z})
(y^{(1)}(\lambda,p_{1})-y^{(2)}(\lambda,p_{1})),
\label{A8}
\end{equation}
\begin{equation}
y^{(1)}=4(y_{11}-2y_{12}+y_{13}),
\label{A9}
\end{equation}
\begin{equation}
y^{(2)}=y_{21}-4y_{22}+y_{23};
\label{A10}
\end{equation}
\begin{equation}
H_{\rm{so}}({\textit{\textbf{k}}}_{1})= \sum\limits_{s=1}^{N}
\sum\limits_{q=1}^{N}{c_{s}}{c_{q}}W(\lambda,p_{1},k_{1z}),
\label{A11}
\end{equation}
\begin{equation}
W(\lambda,p_{1},k_{1z})=t(\lambda,k_{1z})(4w^{(1)}
(\lambda,p_{1},k_{1z})-w^{(2)}(\lambda,p_{1},k_{1z})),
\label{A12}
\end{equation}
\begin{equation}
w^{(1)}=(\lambda p_{1}k_{1z})^{2}
w_{11}+2(2k_{1z}^{2}+p_{1}^{2})w_{12},
\label{A13}
\end{equation}
\begin{equation}
w^{(2)}=(\lambda p_{1}k_{1z})^{2}
w_{21}+(2k_{1z}^{2}+p_{1}^{2})w_{22}.
\label{A14}
\end{equation}

The quantities on the right-hand side of expressions (\ref{A4}), (\ref{A5}), (\ref{A9}),
(\ref{A10}), (\ref{A13}), and (\ref{A14}) are defined as follows:
\begin{eqnarray}
z_{11}=
\sum\limits_{i=1}^{N}\sum\limits_{j=1}^{N}
\frac{\alpha_{1i}\beta_{1i}\,\alpha_{2j}\beta_{2j}}
{(\lambda+\beta_{1i}+\beta_{2j})^{2}}
\,\exp\Bigl(-\frac{\lambda+2\beta_{1i}+2\beta_{2j}}
{\lambda+\beta_{1i}+\beta_{2j}}\,\frac{\lambda p_{1}^{2}}{4}\Bigr),
\label{A15}
\end{eqnarray}
\begin{eqnarray}
z_{12}=
\sum\limits_{i=1}^{N}\sum\limits_{j=1}^{N}\sum\limits_{l=1}^{N}
\frac{\alpha_{1i}\beta_{1i}\,\alpha_{1j}\beta_{1j}\,\alpha_{2l}\,\beta_{2l}}
{(\lambda+\beta_{1ij})^{2}(\lambda+\beta_{1ij}+2\beta_{2l})}
\,\exp\Bigl(-\frac{\beta_{1ij}}{\lambda+\beta_{1ij}}\,
\frac{\lambda p_{1}^{2}}{2}\Bigr),
\label{A16}
\end{eqnarray}
\begin{eqnarray}
z_{21}=
\sum\limits_{i=1}^{N}\sum\limits_{j=1}^{N}\sum\limits_{l=1}^{N}
\frac{\alpha_{1i}\beta_{1i}\,\alpha_{2j}\,\alpha_{2l}\beta_{2jl}}
{(2\lambda+2\beta_{1i}+\beta_{2jl})^{2}}
\,\exp\Bigl(-\frac{\lambda+2\beta_{1i}+\beta_{2jl}}
{2\lambda+2\beta_{1i}+\beta_{2jl}}\,\frac{\lambda p_{1}^{2}}{2}\Bigr),
\label{A17}
\end{eqnarray}
\begin{eqnarray}
z_{22}=
\sum\limits_{i=1}^{N}\sum\limits_{j=1}^{N}
\sum\limits_{l=1}^{N}\sum\limits_{n=1}^{N}
\frac{\alpha_{1i}\beta_{1i}\,\alpha_{1j}\beta_{1j}\,\alpha_{2l}\alpha_{2n}\beta_{2ln}}
{(\lambda+\beta_{1ij})^{2}(\lambda+\beta_{1ij}+\beta_{2ln})}
\,\exp\Bigl(-\frac{\beta_{1ij}}{\lambda+\beta_{1ij}}\,
\frac{\lambda p_{1}^{2}}{2}\Bigr);\nonumber\\
\label{A18}
\end{eqnarray}
\begin{eqnarray}
y_{11}=
\exp\Bigl(-\frac{\lambda p^{2}_{1}}{2}\Bigr)
\sum\limits_{i=1}^{N}\alpha_{2i}\beta_{2i},
\label{A19}
\end{eqnarray}
\begin{eqnarray}
y_{12}=
\sum\limits_{i=1}^{N}\sum\limits_{j=1}^{N}
\frac{\alpha_{1i}\beta_{1i}\,\alpha_{2j}\beta_{2j}}{\lambda+\beta_{1i}+\beta_{2j}}
\,\exp\Bigl(-\frac{\lambda+2\beta_{1i}+2\beta_{2j}}
{\lambda+\beta_{1i}+\beta_{2j}}\,\frac{\lambda p^{2}_{1}}{4}\Bigr),
\label{A20}
\end{eqnarray}
\begin{eqnarray}
y_{13}=
\sum\limits_{i=1}^{N}\sum\limits_{j=1}^{N}\sum\limits_{l=1}^{N}
\frac{\alpha_{1i}\beta_{1i}\,\alpha_{1j}\beta_{1j}\,\alpha_{2l}\beta_{2l}}
{(\lambda+\beta_{1ij})(\lambda+\beta_{1ij}+2\beta_{2l})}
\,\exp\Bigl(-\frac{\beta_{1ij}}{\lambda+\beta_{1ij}}\,
\frac{\lambda p^{2}_{1}}{2}\Bigr),
\label{A21}
\end{eqnarray}
\begin{eqnarray}
y_{21}=
\exp\Bigl(-\frac{\lambda p^{2}_{1}}{2}\Bigr)
\sum\limits_{i=1}^{N}\sum\limits_{j=1}^{N}\alpha_{2i}\beta_{2i}\,\beta_{2ij},
\label{A22}
\end{eqnarray}
\begin{eqnarray}
y_{22}=
\sum\limits_{i=1}^{N}\sum\limits_{j=1}^{N}\sum\limits_{l=1}^{N}
\frac{\alpha_{1i}\beta_{1i}\,\alpha_{2j}\beta_{2j}\,\beta_{2jl}}
{2\lambda+2\beta_{1i}+\beta_{2jl}}
\,\exp\Bigl(-\frac{\lambda+2\beta_{1i}+\beta_{2jl}}{2\lambda+2\beta_{1i}+\beta_{2jl}}\,
\frac{\lambda p^{2}_{1}}{2}\Bigr),\nonumber\\
\label{A23}
\end{eqnarray}
\begin{eqnarray}
y_{23}=
\sum\limits_{i=1}^{N}\sum\limits_{j=1}^{N}
\sum\limits_{l=1}^{N}\sum\limits_{n=1}^{N}
\frac{a_{1i}\beta_{1i}\,a_{1j}\beta_{1j}\,a_{2l}\beta_{2l}\,\beta_{2ln}}
{(\lambda+\beta_{1ij})(\lambda+\beta_{1ij}+\beta_{2ln})}
\,\exp\Bigl(-\frac{\beta_{1ij}}{\lambda+\beta_{1ij}}\,
\frac{\lambda p^{2}_{1}}{2}\Bigr);\nonumber\\
\label{A24}
\end{eqnarray}
\begin{eqnarray}
w_{11}=
\sum\limits_{i=1}^{N}\sum\limits_{j=1}^{N}\sum\limits_{l=1}^{N}
\frac{\alpha_{1i}\beta_{1i}\,\alpha_{1j}\beta_{1j}\,\alpha_{2l}\beta_{2l}}
{(\lambda+\beta_{1ij})^{3}(\lambda+\beta_{1ij}+2\beta_{2l})}
\,\exp\Bigl(-\frac{\beta_{1ij}}{\lambda+\beta_{1ij}}\,
\frac{\lambda p_{1}^{2}}{2}\Bigr),\nonumber\\
\label{A25}
\end{eqnarray}
\begin{eqnarray}
w_{12}=
\sum\limits_{i=1}^{N}\sum\limits_{j=1}^{N}\sum\limits_{l=1}^{N}
\frac{\alpha_{1i}\beta_{1i}\,\alpha_{1j}\beta_{1j}\,\alpha_{2l}\,\beta_{2l}^{2}}
{(\lambda+\beta_{1ij})^{2}(\lambda+\beta_{1ij}+2\beta_{2l})^{2}}
\,\exp\Bigl(-\frac{\beta_{1ij}}{\lambda+\beta_{1ij}}\,
\frac{\lambda p_{1}^{2}}{2}\Bigr),\nonumber\\
\label{A26}
\end{eqnarray}
\begin{eqnarray}
w_{21}=
\sum\limits_{i=1}^{N}\sum\limits_{j=1}^{N}\sum\limits_{l=1}^{N}\sum\limits_{n=1}^{N}
\frac{\alpha_{1i}\beta_{1i}\,\alpha_{1j}\beta_{1j}\,\alpha_{2l}\alpha_{2n}\beta_{2ln}}
{(\lambda+\beta_{1ij})^{3}(\lambda+\beta_{1ij}+\beta_{2ln})}
\,\exp\Bigl(-\frac{\beta_{1ij}}{\lambda+\beta_{1ij}}\,
\frac{\lambda p_{1}^{2}}{2}\Bigr),\nonumber\\
\label{A27}
\end{eqnarray}
\begin{eqnarray}
w_{22}=
\sum\limits_{i=1}^{N}\sum\limits_{j=1}^{N}\sum\limits_{l=1}^{N}\sum\limits_{n=1}^{N}
\frac{\alpha_{1i}\beta_{1i}\,\alpha_{1j}\beta_{1j}\,\alpha_{2l}\alpha_{2n}\beta_{2ln}^{2}}
{(\lambda+\beta_{1ij})^{2}(\lambda+\beta_{1ij}+\beta_{2ln})^{2}}
\,\exp\Bigl(-\frac{\beta_{1ij}}{\lambda+\beta_{1ij}}\,
\frac{\lambda p_{1}^{2}}{2}\Bigr).\nonumber\\
\label{A28}
\end{eqnarray}
In formulas (\ref{A2}), (\ref{A7}), and (\ref{A11}) $\lambda=2/(d_{s}+d_{q})$. The function
$t(\lambda,k_{1z})$ in formulas (\ref{A3}), (\ref{A8}), and (\ref{A12}) has the form
\begin{equation}
t(\lambda,k_{1z})=\pi^{4}\lambda^{3}\exp\left(-\frac{\lambda k_{1z}^{2}}{2}\right).
\label{A29}
\end{equation}
Moreover,
\begin{equation}
\beta_{ijl}=2\beta_{ij}\beta_{il}/(\beta_{ij}+\beta_{il}),\quad
(i\!=\!1,2;\,\,\,j,l\!=\!\overline{1,N}).
\label{A30}
\end{equation}

\setcounter{equation}{0}
\section{\normalsize{Calculation of the profile functions}}
\hspace{\parindent}
The radial parts of the nucleon-nucleus profile functions were calculated in the eikonal approximation
\begin{equation}
\omega_{i}(b_{i})=1-\exp[-\phi_{i}(b_{i})],\quad i\!=\!1,2;
\label{B1}
\end{equation}
where
\begin{equation}
\phi_{i}(b_{i})=-\frac{1}{v}\int_{-\infty}^{\infty}dz\,W\!\!\left(\!\sqrt{b_{i}^{2}+z^{2}}\right)
\label{B2}
\end{equation}
is the scattering phase shift, $v$ is the velocity of the incident nucleon, and $W(r)$ is
the imaginary part of the nucleon-nucleus potential.

Within the framework of the double folding model, the eikonal phase shift can be calculated by
the method described in~\cite{5}.  Let the density distribution of nuclear matter in a nucleon
$(\rho_{i}(r))$ and the amplitude of the nucleon-nucleus interaction in the plane of the impact
parameter $(f(b))$ be described by Gaussian functions
\begin{equation}
\rho_{i}(r)=\rho_{i}(0)\exp(-r^{2}/a_{i}^{2}),
\label{B3}
\end{equation}
\begin{equation}
f(b)=(\pi r_{0}^{2})^{-1}\exp(-b^{2}/r_{0}^{2}),
\label{B4}
\end{equation}
where $\rho_{i}(0)=(a_{i}\sqrt{\pi})^{-3}$, $a_{i}^{2}=r_{0}^{2}=2r_{NN}^{2}/3$,
and $r_{NN}^{2}\cong~0.65~\mathrm{fm}^{2}$ is the mean-square radius of $NN$-interaction.
If the density distribution of the target nucleus (the tabulated distribution~\cite{6} or
the model distribution) is expanded in a series in the Gaussian basis
\begin{equation}
\rho_{T}(r)=\sum_{j=1}^{N}\rho_{Tj}\exp(-r^{2}/a_{Tj}^{2}),\quad
a_{Tj}^{2}=R_{rms}^{2}/j\,,
\label{B5}
\end{equation}
where $R_{rms}$ is the root-mean-square radius of the nucleus, then the formula for the
eikonal phase shift~\cite{5} can be generalized~\cite{4}:
\begin{equation}
\phi_{i}(b_{i})=N_{W}\sqrt{\pi}\,\bar{\sigma}_{NN}\sum_{j=1}^{N}
\frac{\rho_{Tj}\,a_{Tj}^{3}}{a_{Tj}^{2}+2r_{0}^{2}}
\exp\left(-\frac{b_{i}^{2}}{a_{Tj}^{2}+2r_{0}^{2}}\right).
\label{B6}
\end{equation}
Here $N_{W}$ is the normalization factor for the imaginary part of the double folding potential
and $\bar{\sigma}_{NN}$ is the isotopically averaged cross section of the nucleon-nucleus interaction.

Formula (\ref{B6}) was used directly in the calculations of the profile functions (Eqs.~(\ref{B1})),
which were then expanded in the Gaussian basis (formulas (\ref{eq5})).


\end{document}